\documentclass[twoside]{ilcws10}
\usepackage[latin1]{inputenc}
\usepackage[dvips]{graphicx,epsfig,color}
\usepackage{amssymb,amsmath,array}

\def\be{\begin{equation}}
\def\ee{\end{equation}}
\def\bea{\begin{eqnarray}}
\def\eea{\end{eqnarray}}
\def\barr{\begin{array}}
\def\earr{\end{array}}

\def\tev{\; {\rm TeV} }
\def\bea{\begin{eqnarray}}
\def\eea{\end{eqnarray}}
\def\bitem{\begin{itemize}}
\def\eitem{\end{itemize}}

\def\bsub{\begin{subequations}}
\def\esub{\end{subequations}}

\def\f5g{{\cal F}_5^\gamma}
\def\f5z{{\cal F}_5^Z}
\def\h3g{{\cal H}_3^\gamma}
\def\h3z{{\cal H}_3^Z}

\begin{document}
\title{Neutral Trilinear Gauge Boson Couplings in Little Higgs Models}
\author{Sukanta Dutta$^1$, Ashok Goyal$^{2}$  and Mamta$^1$
\thanks{The authors acknowledge partial support from the Department of Science
and Technology, India under grant SR/S2/HEP-12/2006. M would like to acknowledge
 the funds received from 'ILC-India Forum' that made it possible to participate
 in the workshop.}
\vspace{.3cm}\\
1-  Department of Physics, SGTB Khalsa College,  \\
University of Delhi. Delhi-110007, India.
\vspace{.1cm}\\ 
2- Department of Physics \& Astrophysics, University of Delhi \\ 
Delhi-110007, India \\ }
\maketitle
\begin{abstract}
We compute the one loop new physics effects to the CP even triple
neutral gauge boson vertices $\gamma^\star \gamma Z$, $\gamma^\star Z
\,Z$, $Z^\star \,Z\,\gamma$ and $Z^\star\,Z\,Z$ in the context of
Little Higgs models. In addition, we re-examine the
MSSM contribution at the chosen point of SPS1a' and compare with the
SM and Little Higgs models.
\end{abstract}
\vskip -0.5cm
\section{Introduction}
\label{sec:intro}
\vskip -0.3cm
 The measurement of couplings among the electroweak gauge bosons can
 lead to understanding the non-abelian gauge structure of the Standard
 Model (SM) and confront the presence of the new physics above the
 weak scale. The forthcoming experiments at the LHC and proposed ILC
 offer the exciting prospect of probing physics beyond SM. The neutral
 gauge-boson couplings $Z\gamma\gamma$, $ZZ\gamma$ and $ZZZ$ which can
 be studied in $Z\gamma$ and $ZZ$ pair production in $e^+e^-$ and in
 hadron colliders through $e^+e^- \to Z \gamma,ZZ$ and $q \bar q \to
 Z\gamma, ZZ$ respectively have been analyzed within the SM and MSSM
 \cite{Gounaris:1999kf, Choudhury:2000bw}.  In this paper we study the
 CP conserving trilinear neutral gauge-boson couplings in little Higgs
 models and MSSM.

\vskip -0.5cm
\section{Neutral gauge-boson couplings}
\label{ngbc}
\vskip -0.3cm
 Bose-Einstein statistics render the three neutral gauge-boson
 couplings $\gamma\gamma Z$, $\gamma ZZ$ and $ZZZ$ to vanish when all
 the three vector bosons are on shell. The most general CP conserving
 coupling of one off-shell boson ${\rm V}\equiv Z/\, \gamma$ to a pair
 of on-shell $Z \gamma$ and $ZZ$ gauge bosons (all incoming) can be
 written as (see Ref.~\cite{Choudhury:2000bw})
\begin{eqnarray}
\Gamma_{{\rm V} Z\gamma}^{\mu\alpha\beta}(Q,p_1,p_2)&=&i\,\,\left[{\cal H}_3^{\rm V}\epsilon^{\mu\beta\alpha\eta}\,\,{p_2}_\eta+
\frac{{\cal H}_4^{\rm V}}{M_Z^2}\,\,\Big\{\epsilon^{\mu\beta\rho\eta}\,\,{p_2}_{\rho}\,\,{Q}_\eta\,\,Q^{\alpha} \Big\} \right]\\
\Gamma_{{\rm V} ZZ}^{\mu\alpha\beta}(Q,p_1,p_2)&=&i\,\,\Big[{\cal F}_5^{\rm V}\epsilon^{\mu\alpha\beta\sigma}\,\,(p_1-p_2)_{\eta}\Big]
\end{eqnarray}
where $\Gamma_{{\rm V} V_1 V_2}^{\mu\alpha\beta}(Q,p_1,p_2)$ represents
the coupling of off-shell neutral gauge boson $V^\mu$ carrying
momentum $Q$ with the bosons $V_1^\alpha$ and $V_2^\beta$ carrying
momenta $p_1$ and $p_2$ respectively.

 In the SM these couplings vanish at the tree level. These couplings
 can however, be generated at the loop level.  These couplings can in
 general be complex quantities. However, they pick up imaginary
 contribution only when $Q^2$ crosses the threshold for fermion pair
 production (i.e. $Q^2 > 4m_f^2$) for timelike $Q^2$ or when $M_Z^2$
 exceeds this threshold (i.e. $M_Z > 2m_f$) for spacelike $Q^2$.
 Further at the one loop level, the couplings ${\cal
   H}_4^{\gamma}={\cal H}_4^Z = 0$.

The contribution of the fermionic triangle graphs to the trilinear
vector boson couplings ${\cal F}_5^{\gamma ,Z}$ and ${\cal
  H}_3^{\gamma,Z}$ can be expressed in terms of scalar
Passarino-Veltman (PV) functions.
\vskip -0.3cm
\subsection{SM and MSSM Contribution}
\vskip -0.3cm
The contribution to the trilinear neutral gauge couplings in the SM
 arise from the three families of quarks and leptons. The anomaly
 cancellation ensures that all the couplings go to zero for $Q^2$ much
 larger than the fermion pair production threshold. It is obvious that
 of all the thresholds (at $Q^2=4\,\,M_f^2$), the largest contribution
 comes from the heaviest fermion loop. 

The MSSM contribution to the trilinear neutral gauge couplings has
been calculated in the references \cite{Gounaris:1999kf,
  Choudhury:2000bw}.  We re-calculate the MSSM contribution in the
light of the reference point SPS1a' which is defined at a
characteristic scale of 1 TeV with its origin in minimal super-gravity
(mSUGRA) \cite{spa}. The root GUT scale mSUGRA parameters in this
reference point SPS1a' are the gaugino mass $M_{1/2}=250$ GeV, the
universal scalar mass $M_0=70 $ GeV, the trilinear coupling $A_0=-300$
GeV, $\tan\beta({\tilde M})=10$ and sign($\mu)=+1$. Extrapolating
these parameters to $\tilde{M}=1$ TeV generates the MSSM Lagrangian
parameters. The relevant evolved MSSM parameters for our calculations
are the Higgs mixing parameter $\mu=396$ GeV and $M_2=193.2$ GeV.
\vskip -2cm
\subsection{ Contribution in Little Higgs Models}
\vskip -0.3cm
Recently there has been a proposal to consider Higgs fields as pseudo
Nambu-Goldstone bosons of a global symmetry \cite{ArkaniHamed:2002qx,Kaplan:2003uc}
which is spontaneously broken at some high scale.  The realization of
little Higgs mechanism discussed in the literature essentially fall
into two classes  namely the product gorup and
simple group depending on the gauge symmetry. The Littlest Higgs model
(LH)  is a minimal model of the product group
class which accomplishes this task to one loop order within a minimal
matter content. SU(3) simple group model
 is a representative model of the
second class. 

 In the Littlest Higgs model $[SU(2)\times U(1)]^2$ gauge symmetry is
 embedded in an $SU(5)$ global symmetry. The gauge symmetry is broken
 down to the SM $SU(2)\times U(1)$ gauge symmetry by a single vacuum
 condensate $f\approx $ 1 TeV. The new fermionic degrees of freedom in
 the Littlest Higgs model are in the heavy quark sector and consist of
 a pair of vector-like $SU(2)$-singlet quarks that couple to the top
 sector. The resultant top sector consists of a top quark $t$ and its
 heavy partner T whose masses and couplings are given in terms of
 model dependent parameters by $m_t = \frac{\lambda_1\,\,
   \lambda_2}{\sqrt{\lambda_1^2+\lambda_2^2}}$ and $ M_T =
 \sqrt{\lambda_1^2+\lambda_2^2}\,\, f =
 \frac{1}{\sqrt{X_L(1-X_L)}}\,\,\frac{m_t}{v}\,\, f$.
Here $X_L=\lambda_1^2/\,\big(\lambda_1^2+\lambda_2^2\big)$,
$\lambda_1$ and $\lambda_2$ being the couplings that appear in the
heavy quark sector of the interaction lagrangian.
%
Extra Contribution to TNGBC's comes from the triangle
graph with only heavy top and also both the SM top $t$ and heavy top
$T$ simultaneously present in the loop. 

The second class of little Higgs models feature a simple group that
contains an $SU(N)\times U(1)$ gauge symmetry that is broken down to
$SU(2)_L\times U(1)$, giving rise to a set of TeV-scale gauge
bosons. The two gauge couplings of $SU(N)\times U(1)$ are fixed in
terms of two SM gauge couplings, leaving no free parameters in the
gauge sector. Furthermore, due to enlarged $SU(N)$ gauge symmetry, all
fermionic SM representations are extended to transform as fundamental
or conjugate representations of $SU(N)$ giving rise to additional
heavy fermions in all the three quark and lepton sectors. The simplest
realization of this simple group class is the $SU(3)$ simple gauge
model \cite{Kaplan:2003uc} with anomaly-free embedding
of extra fermions. In this model extra fermions present are -- heavy
fermions associated with each $SU(2)_L$ doublet of the SM, new
TeV-scale D and S quarks of charge $-1/3$, heavy third generation
quark T of charge $+2/3$ and electrically Neutral Heavy leptons $N_i$
of three generations. The masses of these heavy
 fermions are given in terms the parameters of the model.
The extra one-loop Contribution to TNGBC's comes
from the triangle graph with mixed SM top $t$ and heavy top $T$, mixed SM
and TeV range quarks of the first two generations and the mixed
neutrino and TeV mass heavy neutrinos $(N_i)$ of all the three
generations in the triangle loop.

Constraints from electro-weak precision measurements require the
breaking scale $f$ to be greater than 5 TeV in the Littlest HIggs
Model and $f>$ 3.9 TeV for $t_{\beta}$ = 3 in the anomaly free
$SU(3)$-simple group~\cite{marandella}. 
 Both
these models  suffer from severe constraints
\cite{Csaki:2003si,Barbieri:2004qk,marandella} from precision
electro-weak measurements. Motivated by these considerations, an
implementation of a discrete symmetry called T-parity is
proposed. T-parity explicitly forbids any tree-level contribution from
the heavy mass states to observables involving only the SM
particles. It also forbids the interactions that impart vev to triplet
Higgs, thereby generating the corrections to precision electro-weak
observables only at the one loop level. In this little Higgs model
with T-parity (LHT) \cite{Cheng:2003ju} there are heavy T-odd partners
of the SM gauge bosons and SM fermions called mirror fermions that
couple vectorially to $Z$.  In the top quark sector, the model
incorporates two heavy T-even and T-odd top quarks in addition to the
T even SM top quark, which are required for canceling the
quadratically divergent contribution of the SM top quark to the Higgs
mass. Further, because of $T$-parity conservation there is no coupling
of $Z_{\mu}$ with a T-odd and and a $T$-even fermion i.e. the coupling
$Z\bar f_+f_- =0$. The $T$-even partner of the top quark $T_+$
however, has both axial and vector couplings with $Z_{\mu}$ and hence
contributes to the triangle loop. In addition to the contribution from
SM fermions in the loop, we now get additional contributions from
heavy $T$-even partner and the top quark as well as contributions from
the triangle loops with mixed contributions from $t$ and $T_+$ quarks.
\begin{table}[!h]
\begin{center}
\begin{tabular}{||c|c|c||c|c|c|c||}
\hline \hline &&&&&&\\[-3mm] 
 {\small {\bf Ratio }}& {\small {\bf $M_{T_+}$( GeV)} }& {\small {\bf $\sqrt{Q^2}$ }} &
 {\small {\bf ${\cal H}_3^\gamma \,(10^{-4})$ }}& {\small {\bf ${\cal H}_3^Z \,(10^{-4})$ }}& {\small {\bf ${\cal F}_5^\gamma \,(10^{-4})$} }& {\small {\bf ${\cal F}_5^Z \,(10^{-4})$ }}\\
\hline\hline & 
 & {\tiny $2m_t $} & {\tiny $-93.40 - \iota \, 0.0149 $} &{\tiny  $ 28.85 + 0 \,\iota \,$ }&
{\tiny $-30.83 +0 \,\iota $ }&{\tiny  $ -20.29 + 0 \,\iota \,$} \\ {\tiny $0.5$}&{\tiny  $ 889.2 $} &
{\bf {\tiny $m_t+M_{T_+} $ }}&  {\tiny $ 4.068 -\iota \, 22.963$ }&  {\tiny  $ -0.705 + \iota \,
7.341$ }&  {\tiny $1.413 -\iota \,7.612 $ }&  {\tiny $ -1.601 - \iota \,10.82$ }\\ & $ $
& {\tiny $2M_{T_+} $} & {\tiny $4.595 -\iota \, 10.66 $} &{\tiny  $ -1.499 + \iota \, 3.875 $}
& {\tiny $ 1.441- \iota \,3.666$ }& {\tiny $ 1.574 -\iota \,6.475$ }\\ 
\hline & $ $ &
 {\tiny $2m_t $ }&  {\tiny $-89.34 - \iota \,0.0130$ }&   {\tiny $ 25.49 + 0\,\iota $ }&  {\tiny $-27.65 +
0\,\iota $ }
&  {\tiny$ -18.06 + 0\, \iota \,$ }\\  {\tiny $1.0$}& {\tiny $711.4 $ }&
  {\tiny $m_t+M_{T_+} $} &  {\tiny $ 0.9154 -\iota \,28.02 $ }& {\tiny $ 5.388 + \iota \,7.730$}
& {\tiny $ -1.313 - \iota \,9.229$ }& {\tiny $ -8.534 - \iota \,11.50$ }\\ & $ $ &
 {\tiny $2M_{T_+} $ }&  {\tiny $3.710 - \iota \,14.24 $ }&  {\tiny $ -0.3113 +\iota \,6.901$} &  {\tiny $
-0.0809 - \iota \,5.481$ }&  {\tiny $ 1.717 -\iota \,10.66$ }\\ 
\hline & $ $ &
{ {\tiny $2m_t $ }}& { {\tiny $ -81.80 - \iota \,0.0094$ }}&{ {\tiny $ 19.84 - \iota \,0.0054$ }}& { {\tiny $
22.00 + 0\, \iota \,$ }}& { {\tiny $-14.31 + \iota \,0.0002 $ }}\\ {\tiny $2.0$}& {\tiny $889.2 $}
& {\tiny $m_t+M_{T_+} $ }& {\tiny $0.7583 - \iota \,19.59 $ }& {\tiny $14.88 + \iota \,3.813
$ }& {\tiny $-4.727 - \iota \,7.226 $ }& {\tiny $ -15.09 - \iota \,8.566$ }\\ & $ $ &
 {\tiny $2M_{T_+} $ }&  {\tiny $1.116 - \iota \,9.099 $} & {\tiny $1.424 + \iota \,9.069$ }&
 {\tiny $-2.866 - \iota \,5.549 $ }&  {\tiny $ -5.549 - \iota \,13.21$ }\\ 
\hline & $ $
& {\tiny $2m_t $ }&{\tiny $ -78.51 - \iota \,0.0079$ }& {\tiny $17.58 -\iota \,0.0068 $ }& {\tiny $
-19.40 + 0 \,\iota $ }& {\tiny $-12.69 + \iota \, 0.0002$ }\\ 
{\tiny $3.0$}& {\tiny $1185.6 $}
&{ {\tiny $m_t+M_{T_+} $ }}& { {\tiny $0.4505 - \iota \,13.03$ }}& { {\tiny $18.70 +\iota \,2.076 $}}
& { {\tiny $-6.278 - \iota \,6.100 $}} & { {\tiny $ -16.71 - \iota \,6.937$ }}\\ & $ $ &
 {\tiny $2M_{T_+} $ }&  {\tiny $-0.623 - \iota \,5.49$ }&  {\tiny $1.759 +\iota \, 9.726$ }&  {\tiny $
-3.972 -\iota \,5.508$ }&  {\tiny $ 4.953 - \iota \,13.55$}\\ 
\hline & $ $ &
  {\tiny $2m_t $ }&  {\tiny  $-77.05 - \iota \,0.0072$} & { {\tiny  $ 16.60 -\iota \,0.0074$ }}&  { {\tiny $
-18.16 + 0\,\iota \,$ }}&   {\tiny $-11.95 + \iota \,0.0002$} \\   {\tiny $4.0$}&  {\tiny $1511.7$ }
&  {\tiny $m_t+M_{T_+} $ }&   {\tiny $ 0.2019 - \iota \,9.182$ }&   {\tiny $ 20.18 +\iota \,1.320$}
&   {\tiny $ -7.086- \iota \,5.541$ }&  {\tiny $ -17.0 - \iota \,6.011$ }\\ & $ $ &
  {\tiny $2M_{T_+} $ }&   {\tiny $-1.712 - \iota \,3.614$ }&   {\tiny $ 1.824 + \iota \,9.939$ }&
  {\tiny $-4.461 - \iota \,5.508 $ }&   {\tiny $ 5.479 - \iota \,13.48$} \\ 
\hline \hline
\end{tabular}
\caption{ \label{tab:lht_rat_var} {\em {\small Ratio, $r = \lambda_1/ \lambda_2$ in
the LHT Model. $f=500$~GeV and $m_t = 175$~GeV.}}}
\end{center}
\end{table}
 \vskip -.5cm
\begin{table}[!h]
\begin{center}
\begin{tabular}{||c||c|c|c|c||}
\hline
\hline
&&&&\\[-3mm]
{\bf $\sqrt{Q^2}$ }& {\bf ${\cal H}_3^\gamma$ }& 
{\bf ${\cal H}_3^Z$ }& {\bf ${\cal F}_5^\gamma$} 
& {\bf ${\cal F}_5^Z$ }\\
(in TeV)&$(10^{-4})$ &$(10^{-4})$&$(10^{-4})$&$(10^{-4})$ \\
\hline\hline
{\small {\bf  $2m_t  $} }&{\small   $-94.17 -\iota \, 0.0158 $} &{\small  $29.53 + 0 \iota\, $ }&
{\small $-31.50+\iota \,0.0149$} & {\small $ -22,.42 +\iota \, 0.0254 $ }\\
\hline
{\small {\bf $m_t+M_T$}} & {\small  {$4.533 -\iota \, 9.136$} }& {\small { $-1.487 + \iota \, 3.008$ }}&  
{\small { $ 1.757- \iota\, 2.802$ }}&{\small { $ -0.1062 -\iota \, 4.751$ }}\\
\hline
{\small {\bf $2 M_T$ }}& {\small   $2.582- \iota\,  3.448 $ }&{\small   $ -2.417 -\iota\, 0.0712 $ }&  
{\small $ 0.5535- \iota\, 0.677$ }& {\small  $ 0.9483 +\iota \, 0.9624$ }\\
\hline
{\small {\bf $M_U$ }}&  {\small {  $3.146 - \iota \,4.660 $ }}&  {\small { $ -2.503 +\iota \, 2.036$ }}&  {\small { 
$1.146 -\iota \,1.152$ }}&  {\small {$ 1.699 - \iota \, 2.449$ }}\\
\hline
{\small {\bf $2M_U$ }}& {\small  $1.372 -\iota \,1.455$ }& {\small  $ 0.1424 - \iota \,1.523$ }&  {\small 
$ 2.947- \iota \,0.191 $ }& {\small  $ -3.151 +\iota \, 2.236$ }\\
\hline
\hline
\end{tabular}
\caption{ \label{tab:su3} {\em {\small The values of various couplings
    (written as complex numbers) at some typical $\sqrt{Q^2}$ (where
    peaks are expected)  in the SU(3) simple model with anomaly free
    embedding. All values correspond to $\tan{\beta} = r = 3$, scale
    $f = 3$TeV and $m_t = 175$GeV. At these values of parameters, the
    mass of heavy top is $M_T = 1.8$~TeV and masses of all other heavy
    fermions have been taken to be $M_i= 3$~TeV.}}}
\end{center}
\end{table}
\vskip -1.cm
\section{Results and Conclusions}
\begin{figure}[tbh!] 
\vskip -5cm
 \begin{center}
  \begin{minipage}[t]{0.33\textwidth}
   \includegraphics[width=5 cm,height=6 cm,]
   {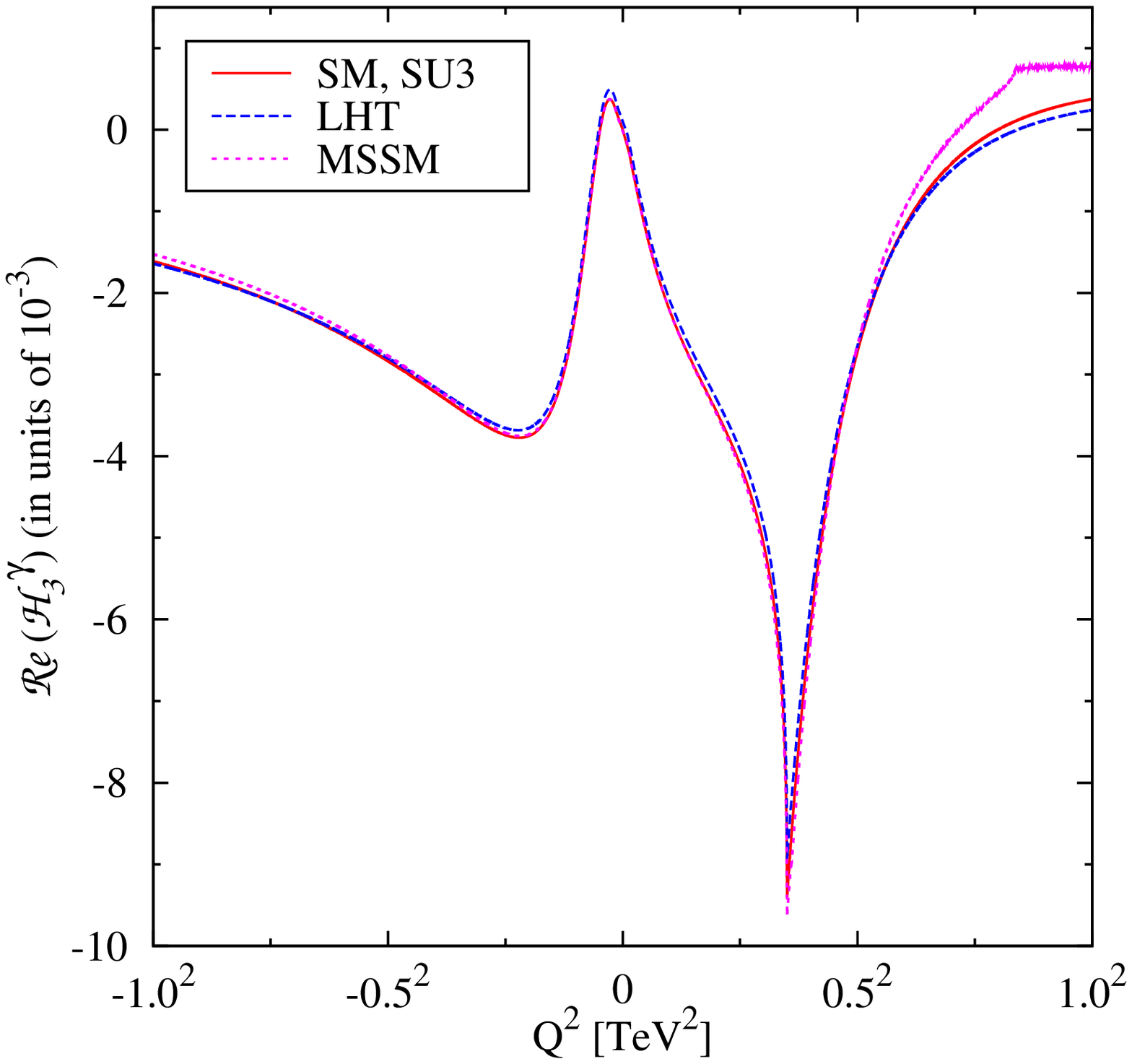}
  \end{minipage}
\hskip 3 cm
  \begin{minipage}[t]{0.33\textwidth}
   \includegraphics[width=5cm,height=6 cm]
   {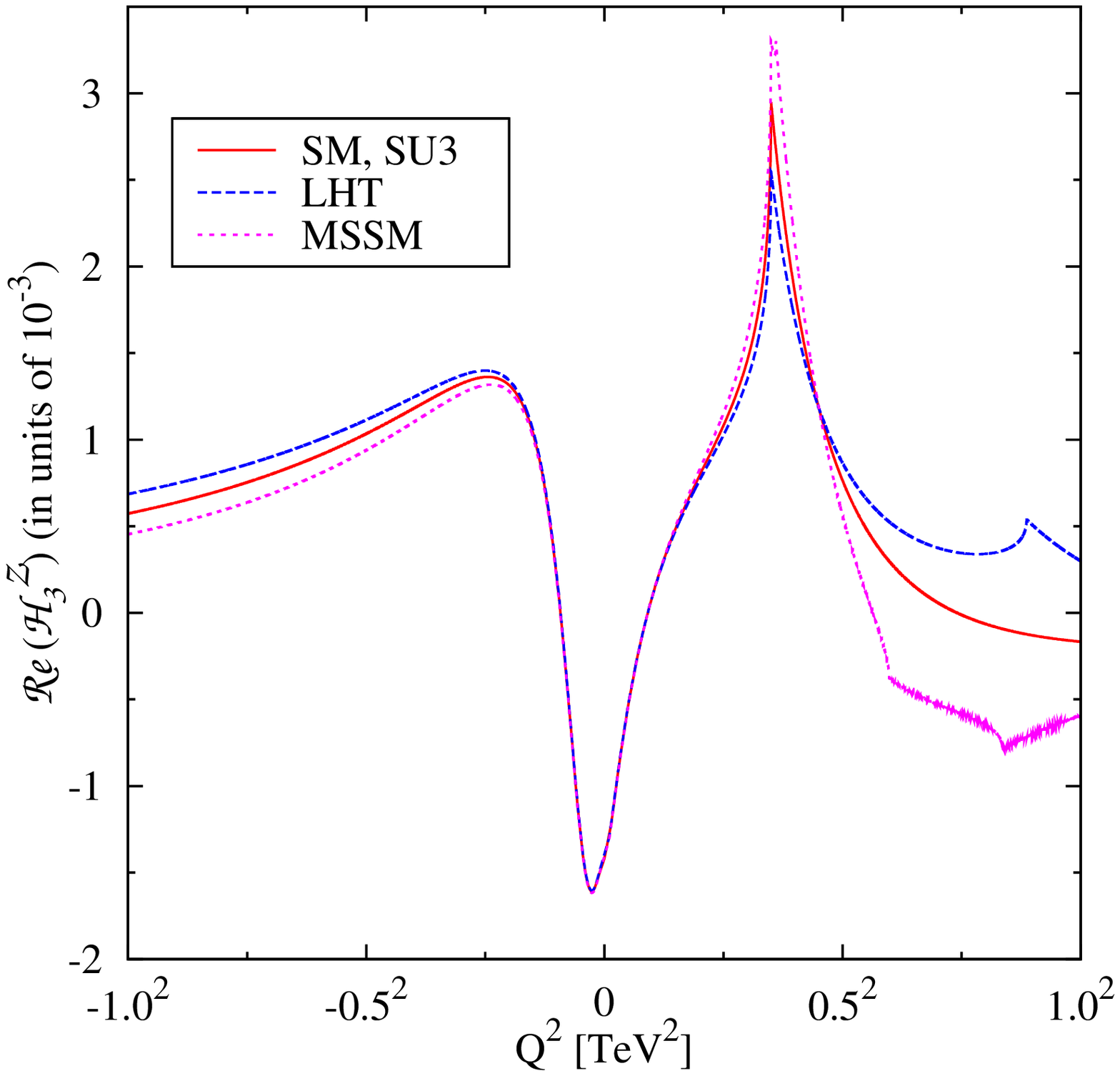}
   \end{minipage}
\vskip -1.5 cm 
\hskip .5 cm
  \begin{minipage}[t]{0.33\textwidth}
   \includegraphics[width=5cm,height=6cm]
   {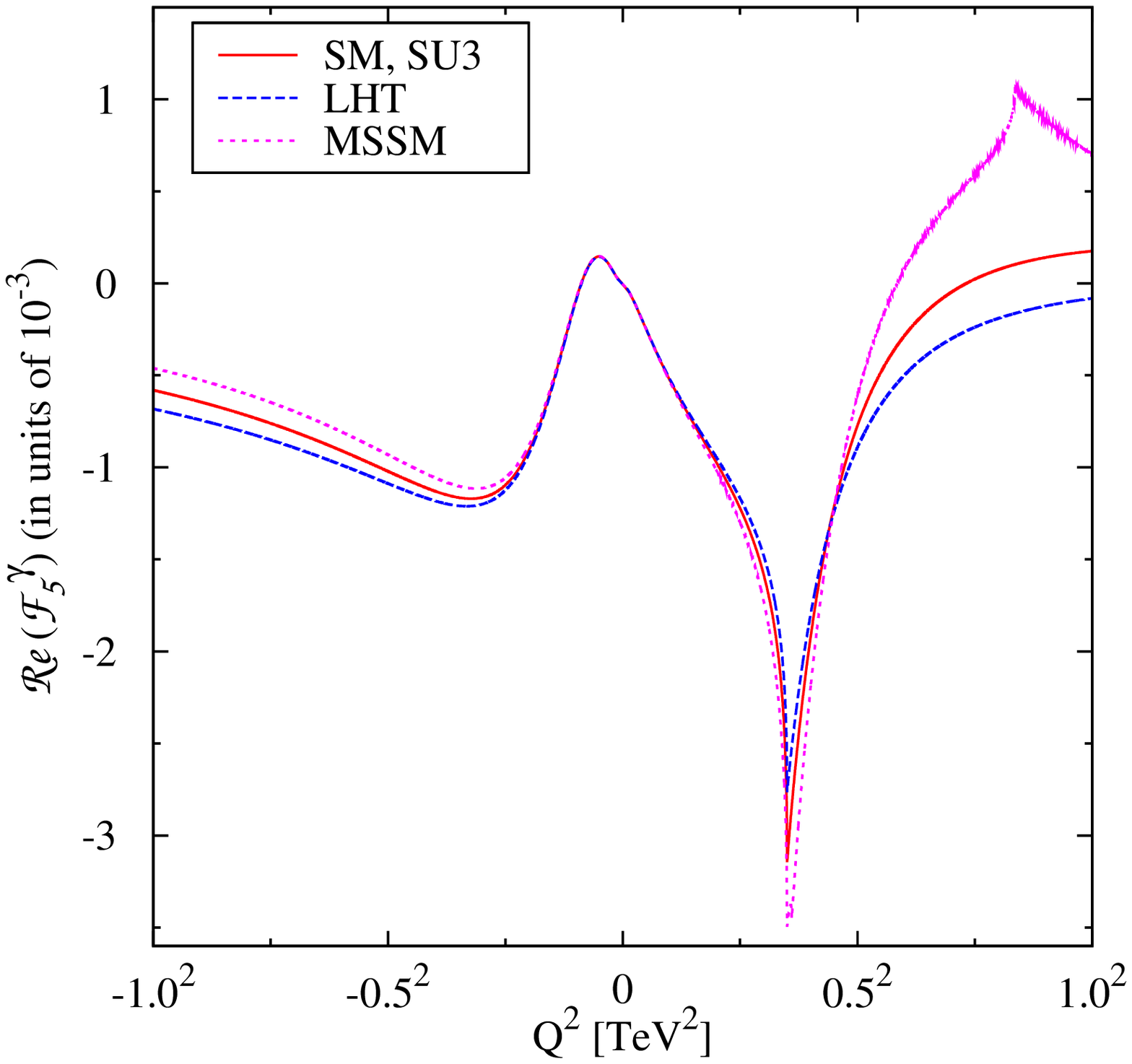}
  \end{minipage}
\hskip 3  cm
  \begin{minipage}[t]{0.33\textwidth}
   \includegraphics[width=5cm,height=6 cm]
   {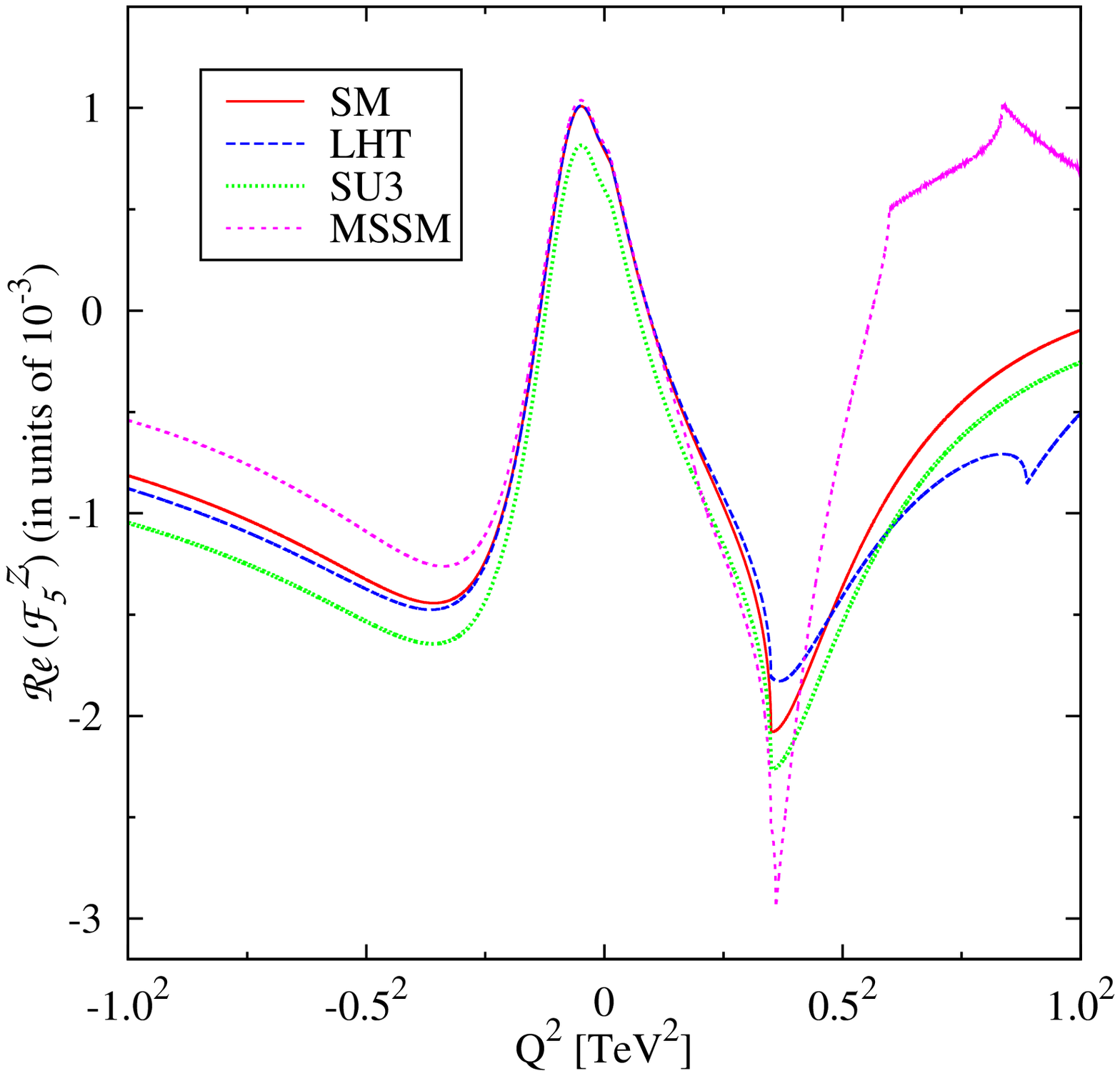}
  \end{minipage}
%
 %
\vskip -1.5cm
  \begin{minipage}[t]{0.33\textwidth}
   \includegraphics[width=5 cm,height=6 cm,]
   {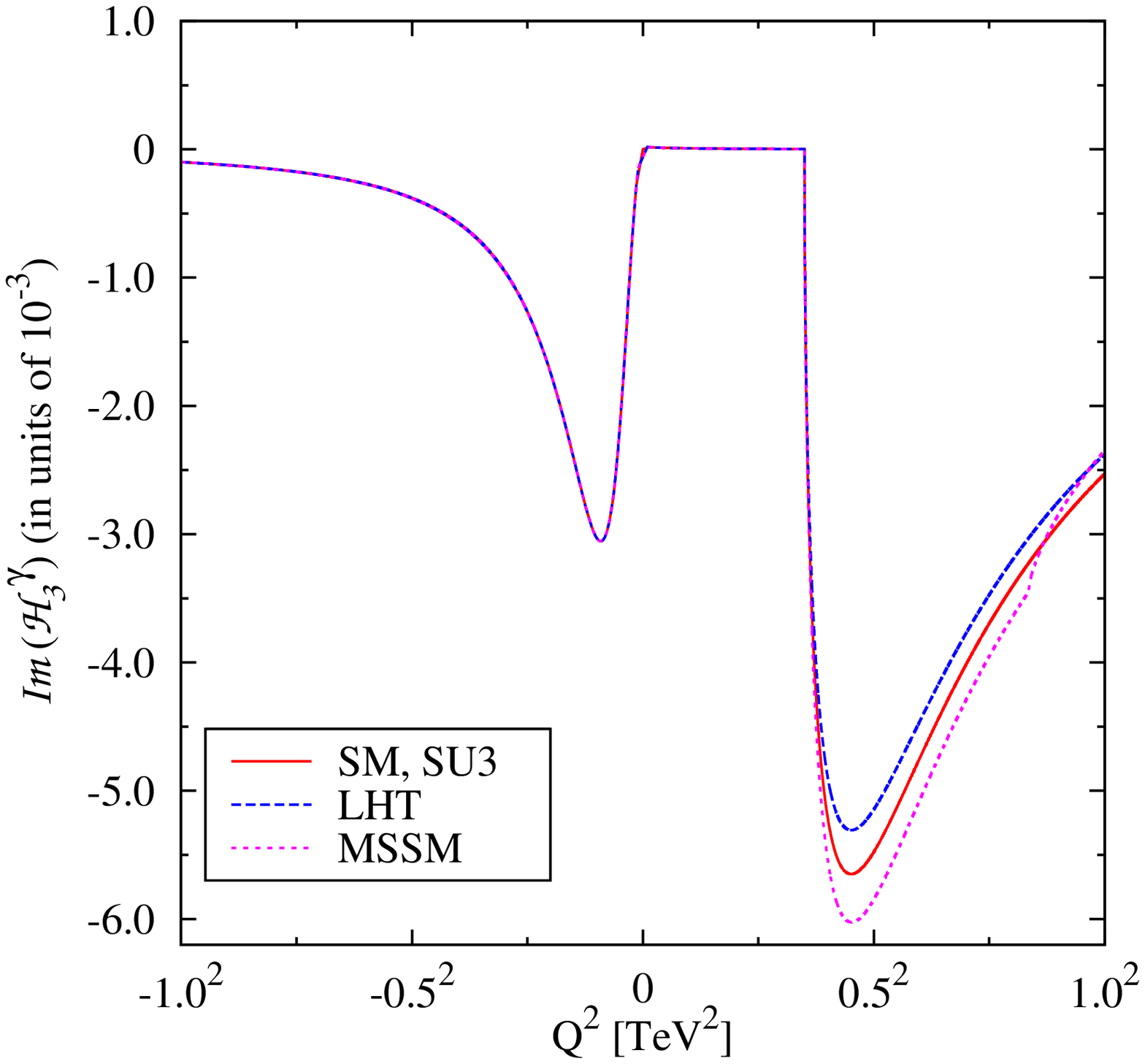}
  \end{minipage}
\hskip 3 cm
  \begin{minipage}[t]{0.33\textwidth}
   \includegraphics[width=5cm,height=6 cm]
   {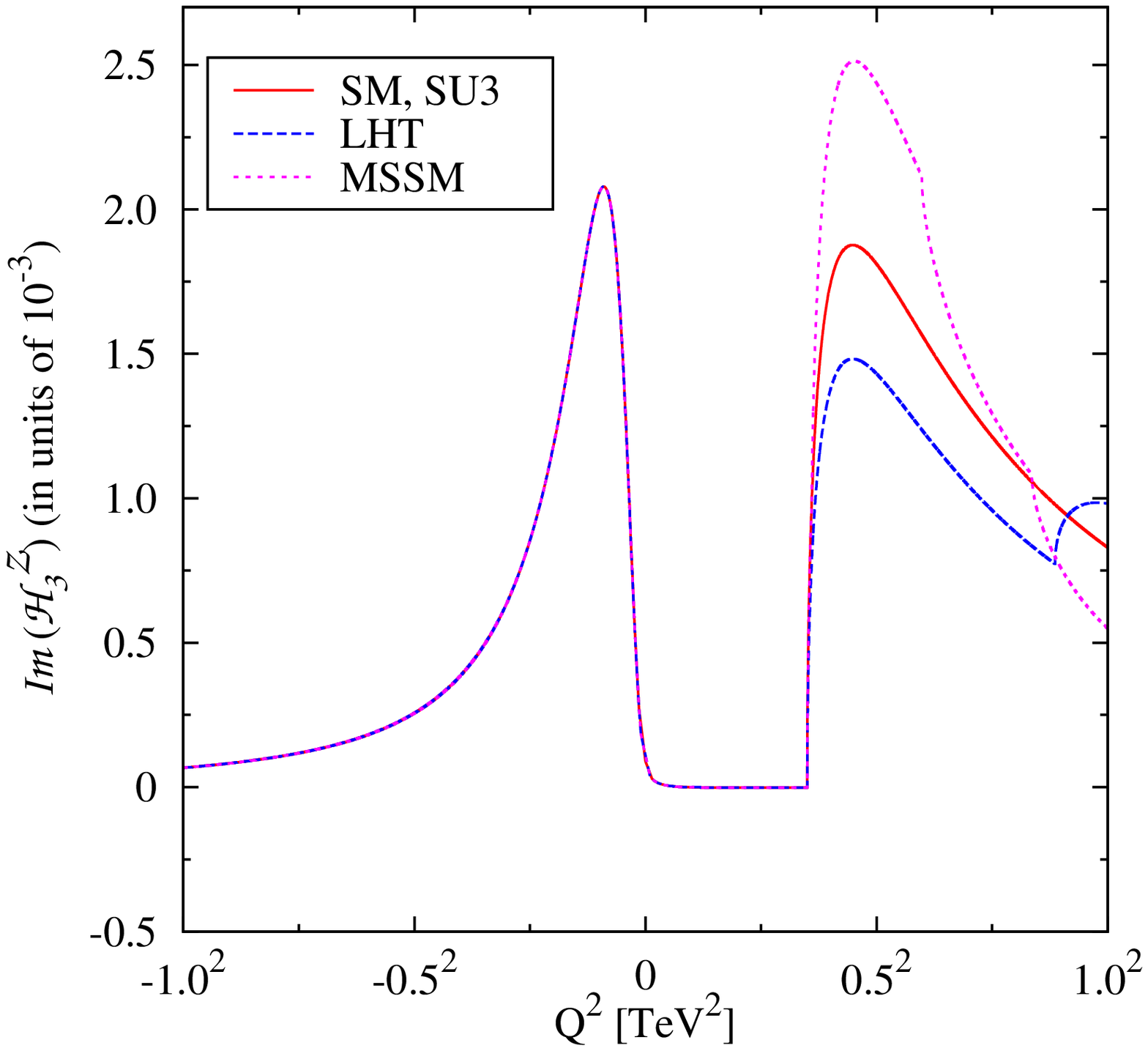}
   \end{minipage}
  \vskip -1.5 cm
\hskip .3 cm
  \begin{minipage}[t]{0.33\textwidth}
   \includegraphics[width=5cm,height=6cm]
   {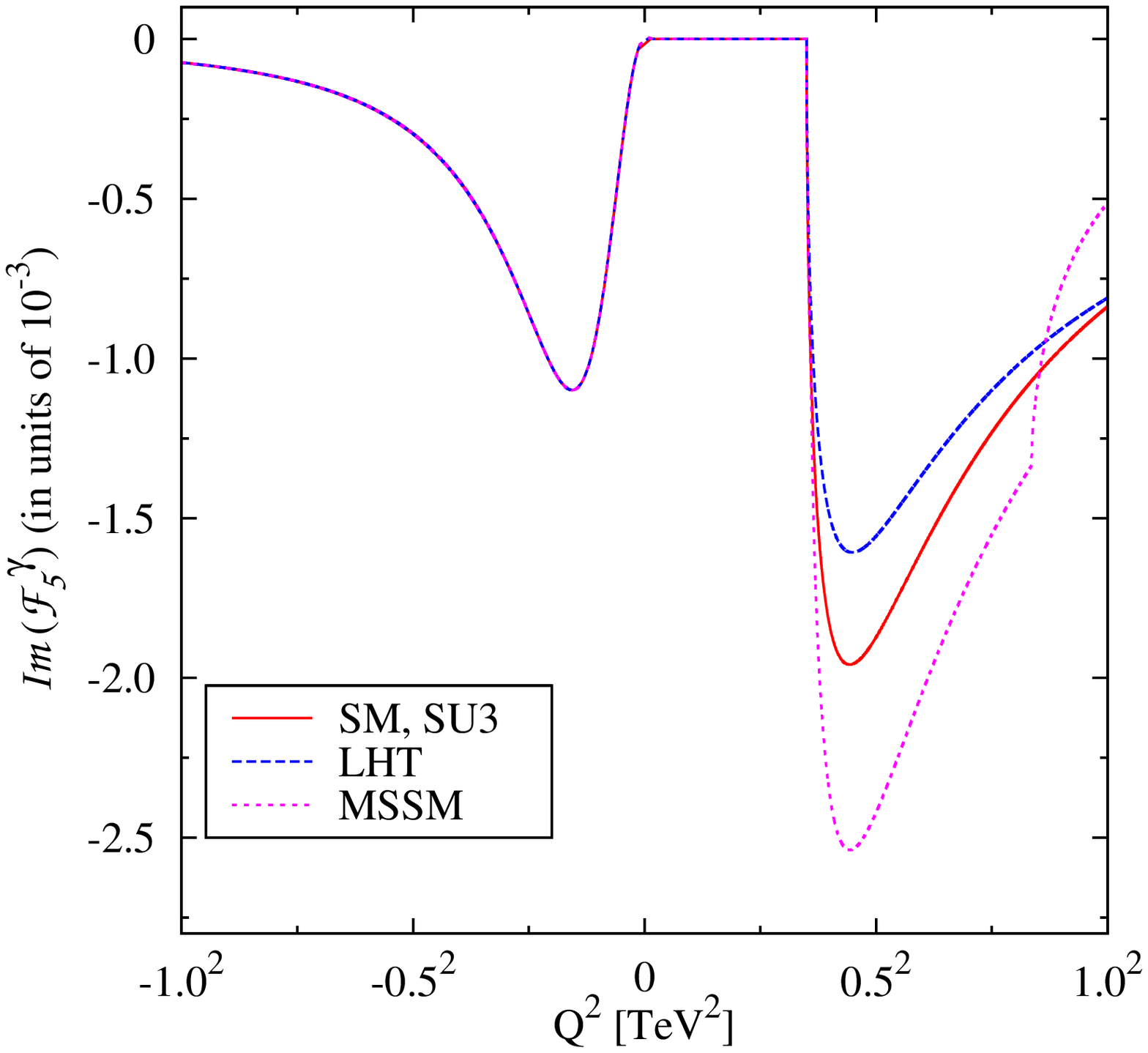}
  \end{minipage}
\hskip 3 cm
  \begin{minipage}[t]{0.33\textwidth}
   \includegraphics[width=5cm,height=6 cm]
   {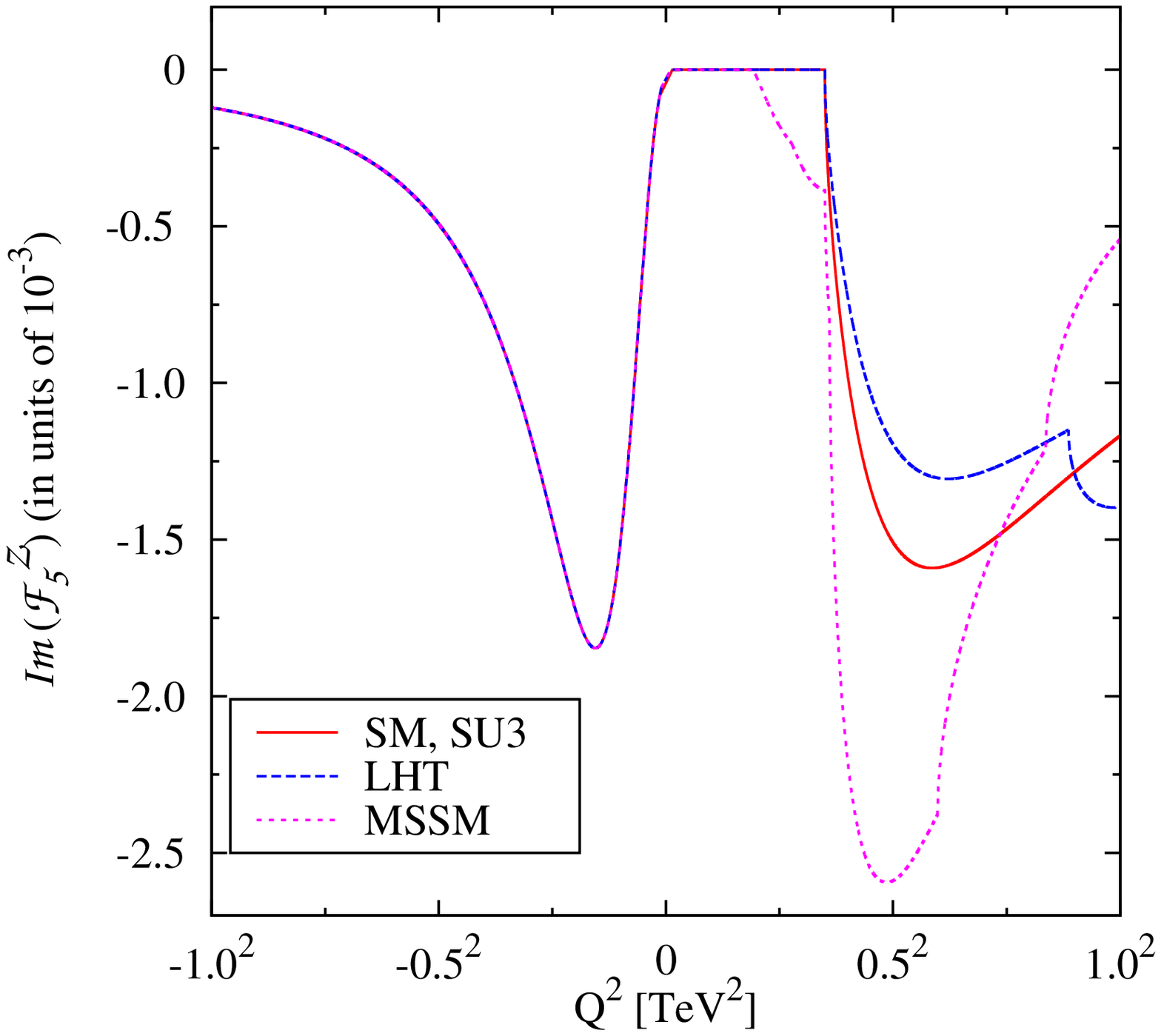}
  \end{minipage}
  \vskip -1.cm
  \caption{\small\em $\sqrt{Q^2}$-variation of the real and imaginary
    parts of the couplings in various models for model parameter
    values $f=500$~GeV, $r=1$ for LHT and $f=3$~TeV, $t_\beta = 3$ and
    masses of all heavy fermions, $M_i = 2$~TeV for the SU(3)
    Model. MSSM parameters are as discussed in the text.}
  \label{CombinedParts}
 \end{center}
\end{figure}

\vskip -0.5cm
We calculate the one-loop contribution to the CP-conserving trilinear
neutral gauge boson couplings in SM, MSSM and the two classes of
Little Higgs Models for various parameters of the models. Certain
features are common to all these graphs which we note here. All
couplings vanish asymptotically for large $\sqrt{Q^2}$ compared to the
highest fermion mass in the theory. The relative importance of the
real and imaginary parts of the couplings is strongly energy
dependent. Below the $2m_t$ threshold, the imaginary parts of all the
couplings are negligible. At and above this threshold the imaginary
parts become comparable or even dominant in comparison to the real
parts. The $\sqrt{Q^2}$ variation of the real and imaginary parts of
the couplings in all the four models is shown in
Fig.~\ref{CombinedParts}. Values at
some typical $\sqrt{Q^2}$ are also given in the
Tables~\ref{tab:lht_rat_var} and \ref{tab:su3} for
different values of parameters of the model.
However, there are certain features that are different in various models.
In the  MSSM  $\sqrt{Q^2}=2\, m_{\chi^+_1}$, which is 
very near to the $2\,m_t$ SM peak resulting in the enhancement of the 
couplings at this point.
This effect is more pronounced in the imaginary parts of the couplings.
In the MSSM  new peaks appear
at $m_{\chi_1^+} +m_{\chi_2^+}\simeq 600$~GeV and  $2 m_{\chi_2^+}\simeq 
800$~GeV which is more pronounced in the real
parts of ${\cal F}_5^\gamma $ and ${\cal F}_5^Z$ . In SM and Little Higgs
Model there is  no such effect upto 1 TeV.

The effect of extra heavy fermions in the LHT Model is to decrease the
threshold effects of the the SM whereas the particles in MSSM enhance
it.
 
The new threshold in the LHT at $\sqrt{Q^2}=m_t+M_{T_+}$ and in the
MSSM as mentioned above are opposite to each other but the magnitudes
are comparable. 
The anomaly free SU(3) simple Model does
not show any appreciably different behaviour than the SM upto $\sqrt{Q^2} =
1 \tev$. 

At higher $\sqrt{Q^2}$, the effect of new
heavy fermions shows up but the threshold values are an order of
magnitude lower than that at the $2m_t$ threshold. However, at these
$\sqrt{Q^2}$, the SM contribution is negligible.

\begin{figure}[!h]
\vskip -1.0cm
\centerline{\includegraphics[width=0.45\columnwidth]{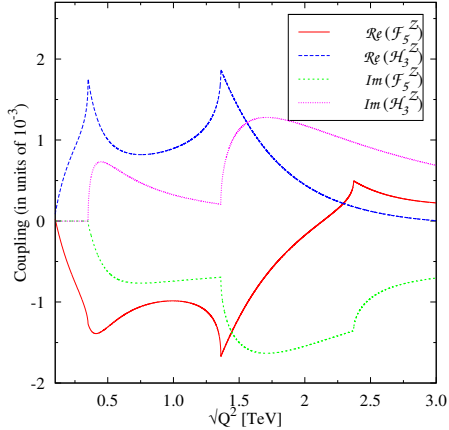}}
\vskip -2.0cm
\caption{\small\em $\sqrt{Q^2}$-variation of real and imaginary parts of
$\h3z$ and $\f5z$ in the range $0-3 \tev$  in Little Higgs Model with T-parity
 for $ r = 3$ and $f=0.5 \tev$. With this
choice of parameters the mass of T-even top, $M_{T_+} = 1.186 \tev$.}
\label{lht_rat3}
\end{figure}

For higher ratios, a very interesting behaviour is shown by the
couplings $\h3z$ and $\f5z$. Not only the
imaginary part becomes appreciable at high $\sqrt{Q^2}$ but also the
threshold values of the couplings at $\sqrt{Q^2} = m_t +
  m_{T_+}$ are higher than those at $\sqrt{Q^2} = 2 m_t$
and are comparable to the SM value (See Figure~\ref{lht_rat3} and
Table~\ref{tab:lht_rat_var}).


Our analysis presented above allows us to confront and discriminate among 
various models considered here on the basis of these couplings.
%

%

\begin{thebibliography}{99}
\small{
%
\vskip -1.5cm
\bibitem{Gounaris:1999kf}
\vskip -.3cm
  G.~J.~Gounaris, J.~Layssac and F.~M.~Renard,
  Phys.\ Rev.\  D {\bf 61}, 073013 (2000)
  [arXiv:hep-ph/9910395].
 %
\bibitem{Choudhury:2000bw}
\vskip -.3cm
  D.~Choudhury, S.~Dutta, S.~Rakshit and S.~Rindani,
  Int.\ J.\ Mod.\ Phys.\  A {\bf 16}, 4891 (2001)
  [arXiv:hep-ph/0011205].
\bibitem{ArkaniHamed:2002qx}
\vskip -.3cm
  N.~Arkani-Hamed, A.~G.~Cohen, E.~Katz, A.~E.~Nelson, T.~Gregoire and J.~G.~Wacker,
  JHEP {\bf 0208}, 021 (2002)
  N.~Arkani-Hamed, A.~G.~Cohen, E.~Katz and A.~E.~Nelson,
  JHEP {\bf 0207}, 034 (2002)
  I.~Low, W.~Skiba and D.~Tucker-Smith,
  Phys.\ Rev.\  D {\bf 66}, 072001 (2002)
  S.~Chang and J.~G.~Wacker,
  Phys.\ Rev.\  D {\bf 69}, 035002 (2004)
  S.~Chang,
  JHEP {\bf 0312}, 057 (2003)
  W.~Skiba and J.~Terning,
  Phys.\ Rev.\  D {\bf 68}, 075001 (2003)
  T.~Han, H.~E.~Logan and L.~T.~Wang,
  JHEP {\bf 0601}, 099 (2006)
\bibitem{Kaplan:2003uc}
\vskip -.3cm
  D.~E.~Kaplan and M.~Schmaltz,
  JHEP {\bf 0310}, 039 (2003)
  [arXiv:hep-ph/0302049];
%
  M.~Schmaltz,
  JHEP {\bf 0408}, 056 (2004)
  [arXiv:hep-ph/0407143];
\bibitem{marandella}
\vskip -.3cm
 G.~Marandella, C.~Schappacher and A.~Strumia,
  Phys.\ Rev.\  D {\bf 72}, 035014 (2005)
  [arXiv:hep-ph/0502096].
\bibitem{Csaki:2003si}
\vskip -.3cm
  C.~Csaki, J.~Hubisz, G.~D.~Kribs, P.~Meade and J.~Terning,
  Phys.\ Rev.\  D {\bf 68}, 035009 (2003)
  [arXiv:hep-ph/0303236];
 \bibitem{Barbieri:2004qk}
\vskip -.3cm
  R.~Barbieri, A.~Pomarol, R.~Rattazzi and A.~Strumia,
  Nucl.\ Phys.\  B {\bf 703}, 127 (2004)
  [arXiv:hep-ph/0405040]; Z.~Han and W.~Skiba,
  Phys.\ Rev.\  D {\bf 72}, 035005 (2005)
  [arXiv:hep-ph/0506206].
\bibitem{Cheng:2003ju}
\vskip -.3cm
  H.~C.~Cheng and I.~Low,
  JHEP {\bf 0309}, 051 (2003)
  [arXiv:hep-ph/0308199].
\bibitem{spa} 
\vskip -.3cm
J. A. Aguilar-Saavedra {\it et. al.}, EPJC (2006) 
 arXiv:hep-ph/0511344V1 
}
\end{thebibliography}
\end{document}